\documentclass[ aps,prl,superscriptaddress,floatfix, twocolumn]{revtex4-2}

\usepackage[utf8]{inputenc} 
\usepackage{amssymb,amsmath}
\usepackage{bbold}
\usepackage{graphicx}
\usepackage{braket}
\usepackage{siunitx}
\usepackage{xcolor}
\usepackage[normalem]{ulem}
\usepackage{notes2bib}
\usepackage{hyperref}
\usepackage{orcidlink}

\def\Rb87{^{87}\mathrm{Rb}} 

\def\ez{\mathbf{e}_z}

\def\HM{\left\{\ket{\text{HM}}\right\}}

\def\OmegaRF{\Omega_{\rm RF}}
\def\OmegaMicro{\Omega_{\mu}}

\begin{document}

\title{Quantum transport in a non-Hermitian 1D synthetic lattice: from quantum Zeno reflection to near-perfect absorption}

\author{Emmanuel.~D.~Mercado-Gutierrez\
\orcidlink{0000-0001-7143-2618}}
\affiliation{Joint Quantum Institute, National Institute of Standards and Technology, and University of Maryland, Gaithersburg, Maryland, 20899-8424, USA}

\author{W.~Xu}
\affiliation{Department of Physics and Astronomy, Purdue University, West Lafayette, IN, 47907, USA}
\affiliation{Purdue Quantum Science and Engineering Institute, Purdue University, West Lafayette, IN, 47907, USA}

\author{E.~Gvozdiovas\
\orcidlink{0000-0003-0782-6705}}
\affiliation{Joint Quantum Institute, National Institute of Standards and Technology, and University of Maryland, Gaithersburg, Maryland, 20899-8424, USA}
\affiliation{Institute of Theoretical Physics and Astronomy, Vilnius University, Saulėtekio Ave. 3, LT-10257 Vilnius, Lithuania}

\author{A.~M.~Piñeiro\
\orcidlink{0000-0001-5988-5227}}
\affiliation{Joint Quantum Institute, National Institute of Standards and Technology, and University of Maryland, Gaithersburg, Maryland, 20899-8424, USA}

\author{Javier~Arg\"uello-Luengo \orcidlink{0000-0001-5627-8907}}
\affiliation{Departament de Física, Universitat Polit\`ecnica de Catalunya, Campus Nord B4-B5, 08034 Barcelona, Spain}
\affiliation{Joint Quantum Institute, National Institute of Standards and Technology, and University of Maryland, Gaithersburg, Maryland, 20899-8424, USA}

\author{W.~D.~Phillips\
\orcidlink{0000-0001-9888-6919}}
\affiliation{Joint Quantum Institute, National Institute of Standards and Technology, and University of Maryland, Gaithersburg, Maryland, 20899-8424, USA}

\author{Q.~Zhou}
\email{zhou753@purdue.edu}
\affiliation{Department of Physics and Astronomy, Purdue University, West Lafayette, IN, 47907, USA}
\affiliation{Purdue Quantum Science and Engineering Institute, Purdue University, West Lafayette, IN, 47907, USA}

\author{I.~B.~Spielman\ \orcidlink{0000-0003-1421-8652}}
\email{ian.spielman@nist.gov}
\affiliation{Joint Quantum Institute, National Institute of Standards and Technology, and University of Maryland, Gaithersburg, Maryland, 20899-8424, USA}

\date{\today}
\begin{abstract}
We experimentally explore the absorption of a propagating wavepacket impinging upon a dissipative region in an engineered quantum system.
We employ a 1D non-Hermitian synthetic lattice with an abrupt interface between dissipative and non-dissipative subchains, using as sites the states of the electronic-ground-state hyperfine manifold in a $\Rb87$ Bose-Einstein condensate.
By tuning the dissipation rate, we observe a progression from ballistic propagation, to near-perfect absorption, to quantum Zeno reflection.
Guided by numerical simulations, we identify that optimal absorption occurs when tunneling and dissipation are properly matched, and find qualitative agreement with an idealized semi‑infinite model across all dissipation regimes.
Our results establish synthetic lattices as a versatile Quantum simulation platform for dissipation-engineered quantum transport and highlight controlled dissipation as a resource for tailoring quantum dynamics.
\end{abstract}

\maketitle

The interaction between quantum systems and their environment often manifests itself as irreversible dynamics, including decoherence, dissipation, and thermalization~\cite{Zurek2003}.
In platforms ranging from photonic systems~\cite{Yariv2002,Chong2010,Xie2011} to analog~\cite{Gross2017} and digital-qubit quantum simulators~\cite{FossFeig2025}, engineered dissipation can give rise to novel phenomena, such as non-Hermitian skin effects~\cite{Hatano1996,Yao2018, Zhou2022, Zhang2022}, exceptional points~\cite{Ding2021}, and $\rm{PT}$-symmetry breaking~\cite{Bender1998, ElGanainy2018}.
Engineered dissipation is also a versatile tool for quantum state preparation and control~\cite{Diehl2008,Daley2014}.
Furthermore, dissipation is fundamental in natural processes such as photosynthesis, where it is important for the transport of energy-carrying excitons and describes their loss by conversion to chemical energy~\cite{Rebentrost2009,Jha2026}.
Situations like this can be modeled as a one dimensional (1D) lattice terminated with a few dissipative sites.
Here we perform a quantum simulation of this model, in which minimizing reflection and maximizing loss is roughly analogous to achieving impedance matching as in transmission lines.

Quantum energy transport and conversion can be studied in ultracold atom arrays by partitioning the system into regions without loss (i.e., the transport region) and with loss (i.e., the conversion zone).
Candidate systems include neutral atoms in optical lattices, tweezer arrays, and chains of trapped ions, where one can apply spatially targeted dissipation using near-resonant focused optical fields~\cite{Maier2019}. 
Efficient energy absorption in the conversion zone is governed by an optimal balance between tunneling and dissipation.
Interestingly, increasing the dissipation away from this optimum decreases absorption by enhancing reflection via the quantum Zeno effect~\cite{Itano1990}. Understanding and controlling this balance is therefore essential for engineering quantum transport.

We implement this transport scenario using a synthetic dimension approach~\cite{Stuhl2015,Kanungo2022}, in which atomic internal states play the role of sites in a 1D chain, allowing site-resolved control over tunneling, dissipation, and lattice geometry beyond what is typically achievable.
(In conventional lattices, avoiding cross-talk between neighboring physical sites requires specialized techniques such as lattice-rearrangement~\cite{Bluvstein2024}, or electron beam microscopy~\cite{Labouvie2016}.)

\begin{figure}[t]
\includegraphics[]{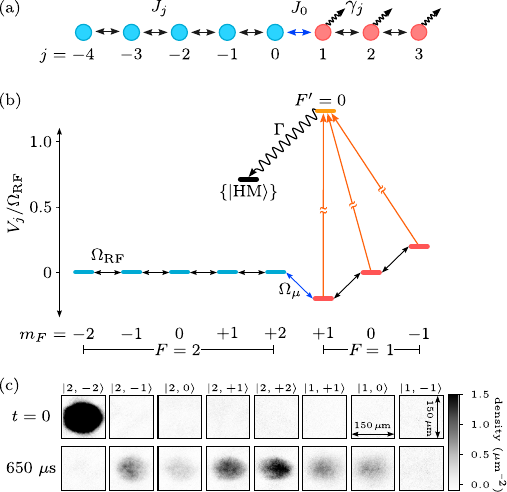} 
\caption{Synthetic lattice. 
(a) Finite chain with non-dissipative (blue, $j\leq 0$) and dissipative (red, $j>0$) subchains.
(b) Experimental realization with the $F=1$ and $F=2$ manifolds of $\Rb87$.
Intra- and inter-manifold couplings are provided by RF and microwave magnetic fields with strengths $\OmegaRF$ and $\OmegaMicro$, respectively.
The $F=1$ ground states are optically coupled to the $F'=0$ excited state; this decays, with rate $\Gamma$, to a range of outgoing momentum states $\HM$.
The potential landscape $V_j$ results from rotating-frame detunings.
(c) Column density associated with each state, measured after evolution times of $t=0$ (top) and $650\ \rm\mu s$ (bottom), with $\OmegaRF = 2\pi\times 1.968(4)\rm\ kHz$, $\OmegaMicro = 2.121(5)\Omega_{\rm RF}$, a microwave detuning $2\pi\times 340(27)\rm\ Hz$, and no dissipation.}
\label{Fig_1}
\end{figure}

This chain is well described by the non-Hermitian tight-binding Hamiltonian
\begin{equation}
\frac{\hat H}{\hbar} \! = \! \sum_{j}\!\Big[\!-\!J_{j}\Big(\ket{j+1}\bra{j} + \mathrm{H.c.}\Big)\!-\!\Big(i\frac{\gamma_j}{2}\!-\!V_{j}\Big)\ket{j}\bra{j}\Big],\label{eq:hamiltonian}
\end{equation}
\noindent
with lattice sites labeled by $j$, on-site energies $V_j$, dissipation rates $\gamma_j$, and nearest neighbor tunneling strengths $J_j$. 
Our system of interest (Fig.~\ref{Fig_1}) is a synthetic lattice, where internal states of the $\Rb87$ atom represent lattice sites.
This lattice is split into two subchains, physically corresponding to the $5\mathrm{S}_{\rm 1/2}$ $F=1$ and $F=2$ hyperfine manifolds.
Figure~\ref{Fig_1}(a-b) illustrates the mapping between the atomic and synthetic degrees of freedom
: the $F=2$, $m_F = -2, \cdots, +2$ states correspond to $j = -4, \cdots, 0$, while $F=1$, $m_F = +1, 0, -1$ maps to $j = 1,2,3$.
The interface is between sites $\ket{j=0}$ and $\ket{j=1}$, where dissipation is non-zero only for sites with $j>0$ (the $F=1$ manifold).
By tuning both the tunneling $J_0$ at the interface between the subchains and the average dissipation $\bar{\gamma}$ (for $j>0$), we identify a regime of optimal absorption and efficient transport when both $J_0\simeq \bar J$ and $\bar{\gamma}\simeq \bar J$, where $\bar J$ is the average tunneling between sites excluding $J_0$.
Moving $\bar\gamma$ away from this condition reduces absorption: as expected for smaller $\bar\gamma$, or from the quantum Zeno effect for larger $\bar\gamma$.

\vspace{3pt}\noindent
\textit{Experimental realization} --- We use a $\Rb87$ Bose-Einstein condensate (BEC) with total atom number $N_{\rm T}\approx10^5$, confined in a crossed optical dipole trap and initially prepared in the $\ket{F=2,m_F=-2}$ ground hyperfine state~\cite{Lin2009b}.
A magnetic field ${\bf B} = B_0 \ez$ splits the Zeeman states within each hyperfine manifold $\{\ket{F, m_F}\}$, allowing a radio-frequency (RF) field to couple adjacent $m_F$ states within those manifolds via the Hamiltonian $\OmegaRF \hat F_x$ (see End Matter), with effective Larmor frequency $\OmegaRF/(2\pi) = 1.968(4)\ {\rm kHz}$. 
This gives an inhomogeneous intra-chain tunneling with average strength $\bar{J} \approx 0.98\OmegaRF$, which we adopt as our unit of energy.
A microwave field couples $\ket{2,2}$ and $\ket{1,1}$ with a tunable Rabi frequency $\OmegaMicro/(2\pi)$, inducing inter-chain tunneling $J_0 = \OmegaMicro/2$.
We induce loss by optically coupling the $F=1$ electronic ground state manifold to the $5\mathrm{P}_{3/2}$ excited level's $F'=0$ state (with lifetime $\Gamma^{-1} = 26.3\ {\rm ns}$).
This results in loss rates $\gamma_j$ that we obtain by fitting the population of BECs prepared in the $F=1$ manifold to a decaying exponential; the resulting values are within about $6\ \%$ of their mean, $\bar\gamma$.
After decaying from the excited state, atoms acquire momentum from the optical recoil imparted by absorption and spontaneous emission.
These high momentum atoms [$\HM$ in Fig.~\ref{Fig_1}(b)] do not return to the BEC, and thereby do not contribute to the observed populations~\cite{Tao2026}. 
The overall rotating-frame energies $V_j$ pictured in Fig.~\ref{Fig_1}(b), computed for nominally resonant coupling fields, include energy offsets due to the slight difference in Landè-$g$ factors between the $F=1$ and $2$ manifolds, as well as the even smaller quadratic Zeeman shifts.

Our experiments all begin with the initial state $\ket{-4}$ (left edge of the lattice) occupied.
At $t=0$, we apply the RF, microwave, and optical fields, thereby initiating tunneling and loss.
Then, after an evolution time $t$, we measure the atom number $N_j(t)$ in each site by rapidly turning off all coupling and trapping fields, applying a Stern-Gerlach pulse~\cite{Lin2009a}, and allowing the Zeeman states to separate for a time-of-flight (TOF).
The density distributions in $F=2$ and $F=1$ are then imaged independently via resonant absorption imaging (after $20\ {\rm ms}$ and $25\ {\rm ms}$ TOF, respectively~\cite{EndMatter}).
Fig.~\ref{Fig_1}(c) shows representative data---including the initial state at $t=0$ (top), and after an evolution time $t=650\ \mu{\rm s}$ (bottom)---from which $N_j(t)$ and therefore occupation probabilities $P_j(t) \equiv N_j(t)/N_{\rm T}$ are determined.

\begin{figure}[t!]
\includegraphics{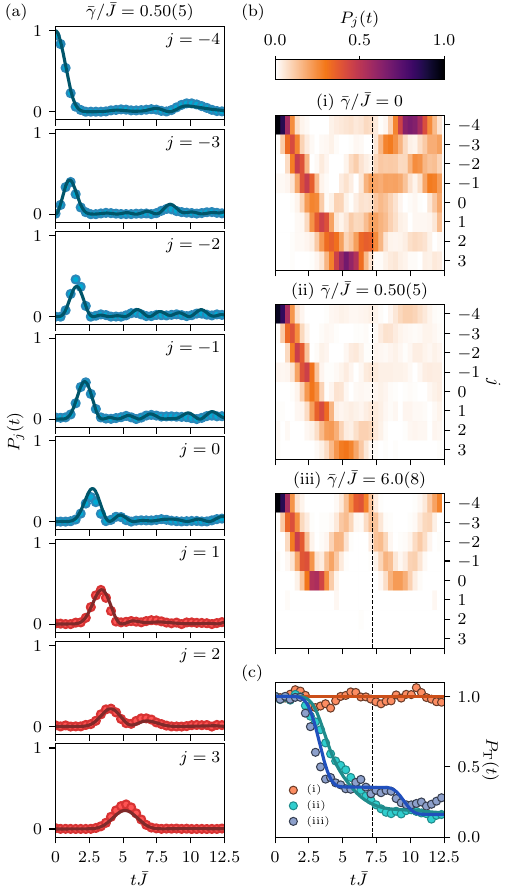}
\caption{
Dynamics. 
(a) Occupation probabilities $P_j(t)$ for dissipative (red) and lossless (blue) sites with $\bar\gamma/\bar{J} = 0.50(5) , J_0/\bar{J} = 1.083(2)$  and $\bar{J} = 2\pi\times1.929(4)\ {\rm kHz}$.
Circles (solid curves) denote experimental data (simulations).
(b) Aggregated experimental data
for three regimes: (i) explicitly lossless [$\bar{\gamma}/\bar{J}=0$], (ii) near-optimal absorption [$\bar{\gamma}/\bar J = 0.50(5)$, identical to (a)], and (iii) quantum Zeno reflection [$\bar{\gamma}/\bar{J}=6.0(8)$]. 
(c) Total probability $P_{\rm T}(t)$ for each regime in (b). The black dashed line at $t\bar{J} = 7.2$ is the dimensionless time selected to measure the single-pass absorption $A$.
}
\label{Fig_2}
\end{figure}

\vspace{3pt}\noindent
\textit{Dynamics} --- We now consider the evolution of the probabilities $P_j(t)$ as the initially localized excitation propagates along the chain.
For example, Fig.~\ref{Fig_2}(a) shows the excitation's temporal wavepacket moving through the lossless subchain (blue) into the dissipative subchain (red) with velocity just over one lattice site per tunneling time $\bar J^{-1}$, and with nearly no change in the temporal width of the wavepacket until encountering the edge of the system at $t \bar J \approx 5$.
The experimental measurements (circles) and numerical simulations with no adjustable parameters (curves) are in good agreement.
Some features present in these data (such as reflections) become more evident when aggregated into images as in Fig.~\ref{Fig_2}(b). Here, panel (ii) is obtained from (a); the trajectory of the primary spatial wavepacket is clearly visible: starting from its initial site, attenuating as it transits the dissipative subchain, and finally returning, greatly attenuated, to its initial site at $t \bar J \approx 10$. 
These data are acquired with parameters [$\bar\gamma/\bar{J} = 0.50(5)$, and $J_0/\bar{J} = 1.083(2)$], which yields near-optimal absorption; when the wavepacket impacts the interface between sites $0$ and $1$ at $t\bar J\approx 3$, it is mostly transmitted with only a small reflection.

Adjusting $\bar\gamma$ away from this value reduces absorption.
In panel (i), decreasing $\bar\gamma$ to zero eliminates absorption as one would expect: the excitation propagates ballistically through both subchains, reflects from the boundary, and returns to its starting point nearly unattenuated, and with some increase in width.
A reflection at the interface is still present, resulting from the small potential offset between sites $j=0$ and $1$ [see Fig.~\ref{Fig_1}(b)].
In panel (iii), making $\bar\gamma$ significantly bigger than $\bar{J}$ reduces absorption by a quantum-Zeno~\cite{Itano1990} enhancement of reflection from the interface and confines the dynamics to the non-dissipative subchain.
This can be understood by interpreting dissipation as an imaginary potential. A step in this potential reflects the wavepacket, much as a real-valued potential step would.

Figure~\ref{Fig_2}(c) plots the total probability $P_{\rm T}(t)=\sum_{j}P_{j}(t)$ (not necessarily conserved in non-Hermitian systems) for the three cases in (b): $\bar\gamma/\bar J =0$ (orange), $\bar\gamma/\bar J=0.50(5)$ (green), and $\bar\gamma/\bar J=6.0(8)$ (blue). 
In all cases, $P_{\rm T}(t)$ remains approximately constant until the wavepacket reaches the interface at $t\bar J\lesssim2$.
When $\bar\gamma>0$, $P_{\rm T}(t)$ then drops as the wavepacket interacts with the dissipative subchain, before temporarily plateauing once the remnants of the wavepacket return to the non-dissipative subchain: from $7 \lesssim t\bar J\lesssim 12$ and $4 \lesssim t\bar J\lesssim 8$ for $\bar\gamma/\bar J\approx 0.5$ and $6.0$, respectively.
At longer times, the wavepacket re-encounters the dissipative subchain, causing $P_{\rm T}(t)$ to decrease further.

\begin{figure}
\includegraphics[]{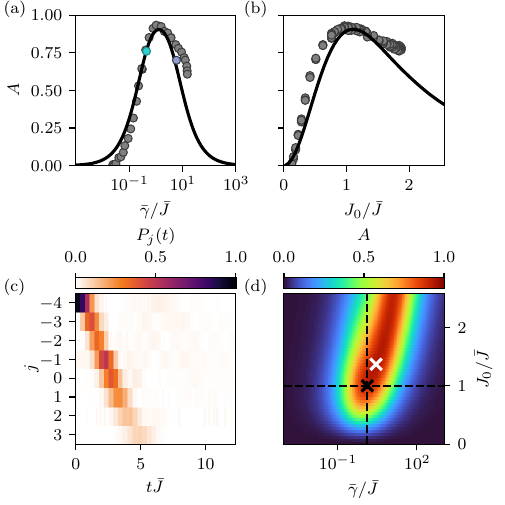}
\caption{
Near-perfect absorption.
(a) Absorption $A$ as a function of $\bar\gamma/\bar{J}$ with $J_0/\bar{J} = 1.083(2)$.
(b) Absorption as a function of $J_0/\bar{J}$ at $\bar\gamma=\bar\gamma_{\rm max} = 1.34(5)\,\bar{J}$.
(c) Wavepacket evolution at $\bar\gamma_{\rm max}$ and $J_{0, {\rm max}}=1.32(7)\,\bar{J}$ found in (a) and (b).
(d) Computed absorption as a function of $\bar\gamma/\bar{J}$ and $J_0/\bar{J}$ with dashed lines indicating the cuts in (a) and (b). 
The white cross marks the global maximum.
}
\label{Fig_3}
\end{figure}

\vspace{3pt}\noindent
\textit{Optimal absorption} --- Motivated by the existence of these plateaus, we use the total remaining probability at $t\bar{J} = 7.2$ [black dashed line, selected to be nominally within both plateau regimes in Fig.~\ref{Fig_2}(c)], to define a proxy $A \equiv 1-P_{\rm T}(7.2/ \bar J)$ for single-pass absorption.
Figure~\ref{Fig_3}(a) plots this proxy as a function of $\bar{\gamma} / \bar J$, including both experimental data (gray symbols) and our model (solid curve); results for the examples in Fig.~\ref{Fig_2}(c) are included as colored circles.
These data confirm the general picture described above: for $\bar{\gamma} / \bar J \ll 1$ absorption is negligible; it increases until reaching a maximum at $\bar{\gamma} / \bar J \approx 1$; absorption then decreases in the $\bar{\gamma} / \bar J \gg 1$ quantum Zeno regime.
Both experiment and theory are consistent with absorption that peaks at $\bar\gamma_{\rm max}/\bar{J} = 1.34(5)$.

Thus far we have focused on tuning the imaginary potential step between $j=0$ and $j=1$.
In Fig.~\ref{Fig_3}(b), we now turn to the effect of the inter-chain tunneling strength $J_0$ with $\bar \gamma = \bar\gamma_{\rm max}$; this includes both experimental data (gray symbols) and our model (solid curve).
For  $J_0/\bar{J} \ll 1$, the two chains are weakly coupled, leading to large reflection and therefore weak absorption; increasing $J_0$ then improves absorption by providing a stronger connection into the dissipative subchain.
For $J_0/\bar J \gg 1$, the two interface sites ($\ket{j=0}$ and $\ket{j=1}$) hybridize, producing local tunneling-induced energy shifts of about $\pm J_0$, the analogue of Rabi splitting in a two-level atom.
This local energy mismatch increases reflection and reduces absorption.
Both experiment and theory are consistent, with the observed maximum absorption at $J_{0, {\rm max}}/\bar{J} = 1.32(7)$, intermediate between these two limits.

With $\bar\gamma_{\rm max}$ and $J_{0, {\rm max}}$ in hand, we return to dynamics; Fig.~\ref{Fig_3}(c) shows more efficient single-pass absorption $A$ than the example in Fig.~\ref{Fig_2}(b)(ii).
Because these parameters are obtained from cross-sectional cuts, they define a constrained optimum rather than the maximum of the full two-parameter landscape.

The computed scan of absorption in Fig.~\ref{Fig_3}(d) shows that the global maximum (white cross) at $\bar{\gamma}/\bar{J} = 2.86$, and $J_{0}/\bar{J}=1.37$ differs slightly from this constrained optimum (dashed lines' intersection).
Nevertheless, the dynamics in Fig.~\ref{Fig_3}(c) support the identification of this cross-sectional optimum as a near-optimal operating point, with predicted absorption $A = 0.90$ close to the global maximum $A = 0.93$. 

The experimentally observed absorption at the constrained optimum is 
$0.94(1)$, comparable to the predicted global maximum; this suggests that simplifications in our model, such as ignoring the slight inhomogeneities in $\gamma_j$ and the attenuation of the loss beam as it traverses the optically thick BEC, led to a ``bank error in our favor''~\footnote{A ``Community Chest'' card in the Hasbro board game Monopoly \textcopyright.  Any mention of equipment, instruments, software, or materials does not imply recommendation or endorsement by the National Institute of Standards and Technology.}.

\begin{figure}
\includegraphics{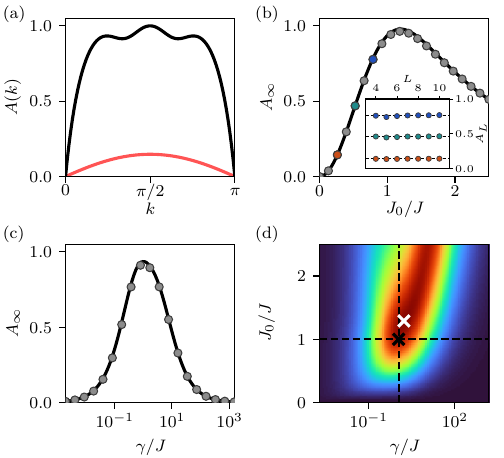}
\caption{
Extended lattice model.
(a) Momentum-resolved absorption $A(k)$ for $\gamma/J=1.14$ (black) and $\gamma/J=50$ (red), with $J_0/J=1$.
(b) Numerical (circles) and analytical (curve) results for $A_\infty$ as a function of $J_0/J$ at $\gamma/J=1.14$; the inset shows the finite-chain  $A$ approaching $A_\infty$ with increasing $L$.
(c) $A_\infty$ as a function of $\gamma/J$ at $J_0/J=1$.
(d) $A_\infty$ as a function of $\gamma/J$ and $J_0/J$ using the same color scale as in Fig.~\ref{Fig_3}(d) with dashed lines indicating the cross sections in (c) and (d).
The white cross marks the global maximum.
}
\label{Fig_4}
\end{figure}

\vspace{3pt}\noindent
\textit{Theoretical extension} --- The observed regimes of absorption and reflection can be understood more generally from a scattering perspective by comparing our finite-chain system to an idealized model with an extended non-dissipative chain, $-L < j\leq 0$, coupled to the same three-site dissipative subchain.
We further simplify the model in Eq.~\eqref{eq:hamiltonian} by setting $V_j=0$, taking $\gamma_j=\gamma$ on the three dissipative sites, and using $J_j=J$ for all intra-chain tunnelings.
The inter-chain tunneling $J_0$, however, remains variable.

We now present an analytic description which is valid in the $L\rightarrow\infty$ limit.
For each incoming wavenumber $k$, the wavefunctions are computed in both the non-dissipative and dissipative subchains, with $\psi_{j\leq0}(k)=e^{ikj}+r(k)e^{-ikj}$ and  $\psi_{j>0}(k) = a_+(k)e^{ik'j}+a_-(k)e^{-ik'j}$, respectively.
Here, $r(k)$ is the reflection coefficient for the probability amplitude, giving the absorption coefficient for the probability $A(k) = 1 - |r(k)|^2$; $a_\pm(k)$ describing the amplitude of forward- and backward-going waves in the dissipative subchain; and the ``wavenumber'' $k'$ is complex~\cite{EndMatter}.

Figure~\ref{Fig_4}(b) shows the absorption coefficient computed for $J_0/J = 1$, comparing a large $\gamma$ case with weak absorption (red) to an intermediate-loss case, $\gamma/J\approx 1$, with complete absorption at $k=\pi/2$ (black).
In all cases, the absorption peaks at $k=\pi/2$ and falls to zero for $k=0$ and $\pi$.
The tight-binding dispersion relation $\omega(k)=-2J\cos(k)$ in the non-dissipative chain has maximum group velocity at $k=\pi/2$, which enhances the penetration into the dissipative subchain and explains the absorption maximum.

We consider transport dynamics starting, as in our experiments, with only the leftmost site occupied (broad in $k$-space).
The absorption of such an initial state can be described in terms of its individual momentum components.  
For example, a uniform distribution across the Brillouin zone would predict the overall absorption $\approx \pi^{-1}\int_0^\pi A(k) \mathrm{d}k$ (neglecting interference terms).
For our lattice, the open boundary at the leftmost site immediately reflects the $-k$ components and imprints a phase $-e^{2ik}$, yielding a weighted average
\begin{align}
A_\infty & \approx \frac{1}{\pi}\int_0^\pi \left[2 \sin^2(k)\right] A(k) \mathrm{d}k \, .\label{eq:reflect_model}
\end{align}
The weight factor amplifies the contribution of $A(k)$ near $\pi/2$ at its maximum; altogether, Eq.~\eqref{eq:reflect_model} provides a remarkably simple analytic model of absorption (see~\cite{EndMatter} for closed expressions).

The curves in Fig.~\ref{Fig_4}(c-d) and the color plot in (e) show the resulting infinite-chain absorption as a function of $\gamma / J$ and $J_0/J$ with optimal absorption of $A_\infty = 0.986$ at $(\gamma,J_0)/J=(1.74,1.29)$; this is in qualitative agreement with the experimental data in Fig.~\ref{Fig_3}(d).
The numerically computed absorption of an $L=5$ finite chain [corresponding to experiment, circles in (c) and (d)] is nearly unchanged from this limit; the inset to panel (c) confirms that overall absorption is remarkably insensitive to system size when $L\geq 4$.
Thus, despite its simplified tunneling and potential landscape, the infinite-chain model supports the same central interpretation as the experiment: absorption is optimized by matching loss and coupling rather than by making the dissipation arbitrarily large.

\vspace{3pt}\noindent
\textit{Outlook and Conclusion}---We used a finite, non-Hermitian, synthetic lattice to realize controllable quantum transport across the regimes of ballistic propagation, near-perfect absorption, and quantum-Zeno reflection.
By independently tuning the loss rate and the interface coupling, we identified an absorption optimum when dissipation and tunneling approximately match. 
The qualitative agreement between finite-chain measurements and an idealized scattering model supports the conclusion that this matching condition captures the same physical tradeoff in both finite-chain dynamics and extended-lattice transport.

Synthetic dimensions provide a natural route to extend this control because tunneling, dissipation, and lattice geometry can be controlled with single-site resolution.
This flexibility could be used to study dissipative transport in higher-dimensional synthetic lattices, in the presence of synthetic gauge fields~\cite{Juzeliunas2008}, or in interacting systems where loss competes with many-body dynamics~\cite{Mak2024}.
Moreover, combining spatial or temporal disorder with dissipation may enable access to phenomena such as dissipation-induced localization~\cite{Patil2015} and Zeno-protected subspaces~\cite{Sun2023}.
More broadly, independently programmable coherent and dissipative couplings make this platform a versatile setting for wave control in non-Hermitian quantum matter.

\begin{acknowledgments}
The authors thank Chenwei Lv for contributing to the theoretical study at an early stage of this project.
This work was partially supported by, the National Institute of Standards and Technology; the National Science Foundation through the Quantum Leap Challenge Institute for Robust Quantum Simulation (grant OMA-2120757); and the Air Force Office of Scientific Research Multidisciplinary University Research Initiative ``RAPSYDY in Q'' (FA9550-22-1-0339).
EG gratefully acknowledges support from the Research Council of Lithuania, Grant No. S-MIP-24-97. 
JAL acknowledges support from the Spanish Ministerio de Ciencia, Innovaci\'on y Universidades (program José Castillejo and grant PID2023-147469NB-C21, financed by MICIU/AEI/10.13039/501100011033 and FEDER-EU) and the Fulbright Program, which is sponsored by the US Department of State and the US–Spain Fulbright Commission.
QZ is supported by the Air Force Office of Scientific Research under
award number FA9550-23-1-0491.
\end{acknowledgments} 

\bibliographystyle{apsrev4-2}
\bibliography{main_2}

\appendix
\section{End Matter}

\vspace{3pt}\noindent
\textit{Intra-chain tunneling inhomogeneity.}---The RF-induced tunneling strengths within each ground-state hyperfine manifold depend on $m_F$.
For the link $\ket{F,m_F}\leftrightarrow\ket{F,m_F+1}$, the tunneling strength is set by the matrix element $\OmegaRF\bra{F,m_F}\hat{F}_x\ket{F,m_F+1}$, where $\hat{F}_x$ is the $x$ component of the angular momentum operator and $\OmegaRF$ is the effective Larmor frequency of the RF drive.
For $F=1$ and $F=2$, the angular momentum operators $\hat{F}^{(1)}_{x}$ and $\hat{F}^{(2)}_{x}$ are
\begin{equation}
    \hat{F}^{(1)}_{x} = \frac{1}{\sqrt{2}}
    \begin{pmatrix}
     0 & 1 & 0\\
     1 & 0 & 1\\
     0 & 1 & 0\\
    \end{pmatrix},
\end{equation}
and
\begin{equation}
    \hat{F}^{(2)}_{x} = 
    \begin{pmatrix}
    0 & 1 & 0 & 0 & 0\\
    1 & 0 & \sqrt{3/2} & 0 & 0\\
    0 & \sqrt{3/2} & 0 & \sqrt{3/2} & 0\\
    0 & 0 & \sqrt{3/2} & 0 & 1\\
    0 & 0 & 0 & 1 & 0\\
    \end{pmatrix}.
\end{equation}
Together, the off-diagonal elements of these matrices describe all tunneling strengths in the synthetic lattice except $J_0$.
The average and standard deviation of the six RF-driven tunneling strengths are $\bar{J}\approx 0.977\,\OmegaRF$ and $\sigma_J \approx 0.212\,\OmegaRF$, respectively.

\vspace{3pt}\noindent
\textit{Dissipation.}---The dissipation rates $\gamma_j$ for the three $\ket{F=1}$ manifold states (i.e., sites $\ket{1}$, $\ket{2}$, and $\ket{3}$) result from coupling to the $\ket{F'=0}$ electronic excited state (with decay rate $\Gamma$) using a resonant optical electric field ${\bf E}$ [see Fig.~\ref{Fig_1}(b)].
We select the polarization of ${\bf E}$ so that the magnitudes of the electric-dipole matrix elements $\bra{F=1,m_F}\hat{\bf d}\ket{F'=0}\cdot {\bf E}$ are nominally independent of $m_F$, and hence $\gamma_j$ is nearly independent of $j$.

\vspace{3pt}\noindent
\textit{Single-realization imaging.}---To image all Zeeman states within a single experimental realization, we sequentially image the Bose–Einstein condensates (BECs) in the $F\!=\!2$ and $F\!=\!1$ hyperfine manifolds.
After traveling through the atoms, the probe light is split into two paths directed onto separate cameras.
Atoms in the $F\!=\!2$ manifold are imaged first using a probe pulse resonant with the $5\mathrm{S}_{1/2}(F\!=\!2)\rightarrow 5\mathrm{P}_{3/2}(F'\!=\!3)$ transition.
Atoms in the $F=1$ manifold are unaffected by this first probe pulse and are then transferred to $F=2$ via a repumper pulse resonant with the $5\mathrm{S}_{1/2}(F=1)\rightarrow 5\mathrm{P}_{3/2}(F'=1)$ transition.
A second probe pulse images these atoms on the second camera \qty{5}{\milli\second} after the initial $F\!=\!2$ image.

\vspace{3pt}\noindent
\textit{Calculation of absorption coefficient.}---In the semi-infinite theory introduced in the main text, the non-dissipative input chain extends over $j\leq0$ and terminates in three dissipative sites, $1\leq j\leq3$.
Away from the interface sites $j=0$ and $1$, the single-particle wavefunction $\psi(j)$ obeys
\begin{equation}
    -J\psi(j+1)-J\psi(j-1)-i\frac{\gamma}{2} \theta(j)\psi(j)=i\partial_t \psi(j),
\end{equation}
where $\theta(j)=1$ for $1\leq j\leq3$ and $\theta(j)=0$ for $j\leq0$.
For an incoming wave with wavenumber $k$, the steady-state wavefunction ansatz is therefore
\begin{equation}
    \psi_k(j)=\begin{cases}
        e^{i k j}+r(k) e^{-ikj}, & j\leq 0,\\
        a_+(k) e^{i k' j}+a_-(k)e^{-i k' j}, & 1\leq j\leq 3\label{ansatz},
    \end{cases}
\end{equation}
where $k'=\arccos[{\cos{k}-i\gamma/(4J)}]$ is a complex wavenumber.
To solve for $r(k)$ analytically, we substitute Eq.~\eqref{ansatz} into the {Schr\"odinger} equation at the interface between $j=0$ and $1$, and into the open boundary condition at the edge of the dissipative chain. As a result,
\begin{align*}
     r(k)&\!=\!e^{i(k+3k')}\frac{J_0^2(1\!+\!2\cos{2k'})\!-\!2J^2e^{ik}(\cos{k'}\!+\!\cos{3k'})}{J^2\!+\!(e^{ik'}\!+\!e^{3ik'}\!+\!e^{5ik'})(J^2e^{ik'}\!-\!J_0^2e^{ik})}.
\end{align*}

When considering transport dynamics starting, for example, from an initial state with a uniform $k$-distribution across $0$ to $\pi$, the overall absorption is $A=1-\sum_{j\leq0}|\pi^{-1}\int_0^\pi r(k)e^{-ikj}\,dk|^2$.
By neglecting the interference terms, we obtain the general form of the approximation used in the main text
\begin{equation}
    A \approx 1-\pi^{-1}\int_0^\pi |r(k)|^2\,dk=\pi^{-1}\int_0^\pi A(k)\,dk.
\end{equation}
Here, the uniform weighting isolates the approximation made by dropping interference terms; the momentum weighting for a specific initial state is not included.

\end{document}